\documentclass[pra,letterpaper,twocolumn,showpacs,superscriptaddress]{revtex4}
\usepackage{graphicx,psfrag,amsmath,amssymb,amsfonts,bbm,latexsym,color,dcolumn}

\begin{document}

\title{Exact solution for the dynamical decoupling of a qubit with telegraph noise}

\author{Joakim Bergli}
\email{jbergli@fys.uio.no}
\affiliation{Physics Department, 
University of Oslo, PO Box 1048 Blindern, 0316 Oslo, Norway}
\author{Lara Faoro}
\email{faoro@physics.rutgers.edu}
\affiliation{Department of Physics and Astronomy, Rutgers University,
136 Frelinghuysen Road, Piscataway 08854, New Jersey, USA}

\date{\today}

\begin{abstract}
We study the dissipative dynamics of a qubit that is afflicted by classical
random telegraph noise and it is subject to dynamical decoupling. 
We derive exact formulas for the qubit dynamics at arbitrary working points 
in the limit of infinitely strong control pulses (bang-bang control) and we 
investigate in great detail the efficiency of the dynamical decoupling 
techniques both for Gaussian and non-Gaussian (slow) noise at qubit 
pure dephasing and at optimal point. We demonstrate that control 
sequences can be successfully implemented as diagnostic tools 
to infer spectral proprieties of a few fluctuators interacting with the qubit.
The analysis is extended in order to include the effect of noise in the 
pulses and we give upper bounds on the noise levels that can be tolerated 
in the pulses while still achieving efficient dynamical decoupling 
performance. 
\end{abstract}


\maketitle

\section{Introduction}
Small superconducting circuits containing Josephson junctions have received a great deal of attention recently as promising candidates for 
scalable quantum bits (qubits). Unfortunately, despite the 
last few years of remarkable experimental and theoretical breakthroughs
\cite{Nakamura1999,Vion2002,Chiorescu2003,Wallraff2004,Martinis2002}, 
neither of these circuits satisfies the requirement that
the  coherence times must be  considerably longer 
than the gate operation times \cite{DiVincenzoCriteria}. Thus a  
major challenge for the next years is to achieve  control of
the decoherence afflicting these devices. 

It is generally recognized that in superconducting qubits the main 
sources of noise are: the external 
circuitry, the motion of trapped vortices in the superconductors and the 
motion of background charges (BCs) in associated dielectrics and oxides. 
While the noise originating from the external circuitry can be reduced and 
practically eliminated by improving the circuit design and the motion of 
trapped vortices can be suppressed by making the superconductor films sufficiently narrow, it is currently not known how to suppress and 
control charge noise. Indeed, moving BCs of {\em unknown} origin
are clearly responsible for the 
relatively short coherence time in the smaller Josephson charge qubits 
\cite{Nakamura1999}. In addition, BCs limit the performances of the 
flux qubits \cite{Chiorescu2003},
the phase qubits \cite{Martinis2002} and the hybrid charge-flux qubits 
\cite{Vion2002} because they induce fluctuations in the critical current 
of the Josephson junctions and lead to qubit dephasing. 

It is firmly established now that in all superconducting qubits both 
the spectrum of charge noise and critical current fluctuations display a
$1/f$ behavior at low frequencies. Specifically, charge noise spectra ${S_{q}(f)=\frac{\alpha ^{2}}{f}}$ have been directly measured up to frequencies ${f\lesssim10^{3}\;}$Hz  \cite{Zimmerli1992,Visscher1995,Zorin1996,Vion2002,Nakamura1999,Martinis2002}. The intensity of the noise varies in the small range 
${\alpha =10^{-3}-10^{-4}e}$ and because its magnitude is greatly reduced by the echo techniques \cite{Nakamura2002,Ithier05}, it is commonly believed that $1/f$ noise does not extend 
above $f_{m}\sim 1\; $MHz. Similarly, it has been shown that 
critical current fluctuations display a $1/f$ behavior at 
${f \lesssim 10 \;}$Hz  although very little is known at 
higher frequency  \cite{Wellstood1998,Wellstood2004,VanHarlingen2005}. 

Both charge noise and critical current fluctuations 
can be phenomenologically explained by modeling the environment as a 
collection of {\em discrete} bistable fluctuators, representing 
charged impurities hopping between different locations in the substrate or 
in the tunnel barrier. 
Because the hopping time is exponential in the height of the barrier 
separating different impurity states,  
one expects a wide distribution of the relaxation times for the fluctuators.
Such wide distributions naturally lead to the low frequency 
$1/f$ dependence. 

The discrete nature of the environment 
has important consequences for the qubit dephasing. 
As was shown in Ref.~\onlinecite{Paladino2002}, the effect of a 
bath of fluctuators on the qubit is very different from that of a continuum of 
linearly coupled oscillator modes, which is the 
subject of the Caldeira-Leggett dissipation theory \cite{Leggett87}. 
The discreteness of the environment becomes especially important 
for the slow modes which switching rates that are longer or comparable to 
the operational time of the qubits. 
Qubit interaction with slow fluctuators 
might lead 
to non-Markovian errors that are difficult to correct \cite{Alfieri} and to 
a fast initial qubit dephasing \cite{Falci2005}.

It has been recently suggested that 
dynamical decoupling techniques \cite{Viola1998} could be very
promising strategies for suppressing the noise generated by 
slow fluctuators in superconducting qubits 
\cite{Shiokawa2002,Gutmann2003,Faorodec,Falcibb}.
Unfortunately, because the behavior of the 
charge noise power spectrum is known only in very limited interval of 
frequencies, it is still not obvious whether it is possible to develop 
{\em efficient} dynamical decoupling schemes for these devices.

Indeed, besides the $1/f$ low frequency part of the charge noise 
spectrum, so far it is only known 
that charge noise displays an ohmic behavior at very high frequencies 
($7 \;\text{GHz} <f<100 \;\text{GHz}$) \cite{Astafiev2004}. 
The noise spectrum at intermediate frequencies can only be inferred from 
its effects on the qubit and so its detailed behavior is not known. 
This missing part is very important for the development of dynamical 
decoupling techniques, 
because the presence of noise at the time scales corresponding to the 
time interval between echo pulses limits the error suppression capability of
these techniques.

Interestingly, it has been argued that dynamical decoupling techniques might be
also used to extract information on this missing part of the noise spectrum, as a sort of {\em diagnostic tools} to infer spectral proprieties of the charge fluctuators coupled to the qubit\cite{Faorodec,Falcibb}. 
Several numerical results pointed out that
the interplay between the applied pulses 
and the qubit dynamics might originate a rich behavior in 
the dissipative dynamics depending on the characteristics of the fluctuators 
and the working point of the qubit. This conjecture is supported also by 
a few analytical results.
For example, in Ref.~\onlinecite{Faorodec} the authors derived 
the analytical solution for the ideal (i.e. with no errors in 
the pulses) dynamical decoupling of a qubit coupled to many fluctuators at
pure dephasing, while in Ref.~\onlinecite{Falcibb} the analytical solution 
for the ideal decoupling of a qubit coupled to a fluctuator at a generic 
working point has been sketched.
However, we believe that from these analytic results it remains still 
difficult to estimate the {\em effective} diagnostic capabilities of 
the dynamical decoupling techniques. Moreover, 
in a realistic set-up, one should  expect that the dynamical 
decoupling is not ideal and it should be desirable to extend this analysis 
to include errors in the pulses.
 
This paper aims (i) to simplify the previous analytical solutions of the 
dynamical decoupling at pure dephasing, (ii) to find the analytic solution
of the dissipative qubit dynamics at the optimal working point, (iii) 
to extend these analytical results to the case of non-ideal control pulses.
We hope that this novel approach and our analytical solutions 
might be relevant in the design and interpretation of new experiments. 

A remark is needed at this point. In this paper we shall study only the 
dissipative dynamics of the qubit coupled to a {\em single 
classical} fluctuator that is subject to dynamical decoupling. 
There are several reasons for this choice. 
First, there are interesting 
experimental situations where there is evidence of a single  fluctuator 
dominating the qubit dynamics \cite{Nakamura2002,Esteve,Galperin2004}. 
In this case,
the analytical solutions derived in the following sections are 
sufficient to interpret the experimental data. 
Second, we believe that even from this simple solvable model one can obtain
some important insights and intuitions on the slow frequency noise and 
especially on its non-Gaussian nature. Third, at 
pure dephasing, it is trivial to extend our analytical results to many 
fluctuators interacting with the qubit, because the total response is simply  
the linear combination of the contributions from each individual 
fluctuator \cite{Galperin2004}.

On the other hand, the fact that we are considering a classical fluctuator 
implies that our discussion most likely will be relevant to 
superconducting qubits made out of a Cooper pair box or more generally 
to those superconducting circuits containing ultrasmall junctions (having 
area ${\lesssim 0.05 \mu \text{m}^2}$) where the presence of few dominant 
classical fluctuators have been clearly observed in critical current 
fluctuations \cite{Wakai87,Rogers85}. 

However, it is relevant to mention here that also experiments in phase qubits  
\cite{Martinis2002,Simmonds2004} and very recently in flux qubits 
\cite{Nakamura2006} show the presence of single charged impurities strongly 
coupled to the qubit and seriously affecting its dynamics. It would be very interesting to see whether dynamical decoupling sequences could help the characterization of the noise even in these cases. However, since 
it is reasonable to expect that those charged impurities are 
{\em quantum} Two Level Systems present in the insulating barrier of the 
Josephson junctions \cite{Faoro2005}, the analysis discussed here should 
be extended in order to take into account the {\em quantum} nature of the fluctuator. 
This will be the subject of future work.

This paper is organized as follows. In Sec. \ref{qubit} we shall briefly 
recall the physics of the Cooper pair box, we shall derive the total 
Hamiltonian of the qubit in interaction with a bath of fluctuators and 
with applied an external control field and we shall finally explain how 
sequences of pulses can be generated in this device.  
In Sec. \ref{ideal} we shall present analytical results for the dynamics 
of the qubit subjected to {\em ideal} dynamical decoupling 
(i.e. with no errors in the pulses) at different qubit working points. 
Specifically, in Subsec. \ref{idealpd} we shall consider the qubit at 
pure dephasing and we shall greatly simplify the analytical result 
given in Ref.~\onlinecite{Faorodec}, thereby immediately elucidating 
the efficacy of the dynamical decoupling as a 
diagnostic tool. In Subsec. \ref{idealop} we shall provide the 
analytical solution for the dissipative dynamics of the qubit subjected 
to dynamical decoupling at the optimal point. This result will allow 
us to comment on the evidence of the anti-Zeno effect in this regime 
\cite{Faorodec,Falcibb} and to explain how the latter can 
be conveniently avoided. 
In Sec. \ref{imperfect} we shall discuss dynamical decoupling sequences 
with imperfect pulses. The consequences of imperfect bang bang pulses 
have been already discussed, for the case of a single fluctuator, in 
Ref.~ \onlinecite{Gutmann2004}. Here, by using our approach, we are able to
derive bounds on the pulse noise that can be tolerated by 
the device while still achieving dynamical decoupling. 
Finally, Conclusions are discussed in Sec. \ref{conc}. 

\section{Cooper pair box, charge noise and control}
\label{qubit}
\begin{figure}[h]
\begin{center}
\includegraphics[width=5.0cm]{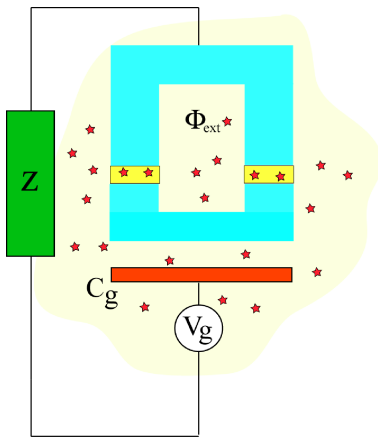}
\end{center}
\caption{Sketch of a Cooper pair box afflicted by charge noise present 
in the amorphous materials of the substrate and the junction barriers).  
\label{cpbfig}}
\end{figure}

A Cooper pair box \cite{Nakamura1999,Vion2002} is a small 
superconducting island connected to a reservoir via two Josephson junctions (see Fig.~\ref{cpbfig}).
The box is characterized by two energies: the Josephson coupling energy  
$\displaystyle{E_J\left (\Phi_{ext} \right )=\sqrt{E_{J1}^2 + E_{J2}^2 + 2 E_{J1} E_{J2} \cos \left (
\pi \Phi_{ext}/\Phi_0 \right )}}$ and the charging energy $E_c=e^2/2C$. 
${E_{Ji}}$, $i=1,2$ are the Josephson energies and ${\Phi_{ext}}$, ${\Phi_0}$ 
are respectively the external magnetic flux 
piercing the superconducting loop and the flux quantum.  
$C$ is the total capacitance of the box and $e$ denotes the electron charge.

Under the assumptions that (i) the charging energy $E_c$ is much larger than
the Josephson coupling $E_J$, (ii) the superconducting gap 
$\Delta$ of the box is larger than its charging energy 
(so to avoid quasi-particle excitations), (iii) the temperature $T$ is small 
compared to all these energy scales,
it is possible to describe the dynamics of the Cooper pair box as a two 
state system, whose Hamiltonian in the charge basis $\{ |0 \rangle, |1 \rangle \}$ reads:
\begin{equation}
H_S=\frac{\delta E_c}{2} \sigma_z +\frac{E_J (\Phi_{ext})}{2} \sigma_x\;
\label{Cpb}
\end{equation}
where $\sigma_{x,z}$ are the  
Pauli matrices. Both the bias ${\delta E_c=E_c (1-C_g V_g/e)}$ and the 
Josephson energy can be easily tuned by varying, respectively, the applied 
gate voltage $V_g$ and the magnetic field $\Phi_{ext}$ threading the loop 
containing the island.

The Hamiltonian given in Eq. (\ref{Cpb}) describes the dynamics of
an {\em ideal} Josephson charge qubit. However, it is known that
many charged impurities are present in the amorphous materials both of 
the substrate (on which the Cooper pair box is grown) and of the 
Josephson barriers. These charges, by interacting with the qubit, are 
responsible for its decoherence.

In order to write an Hamiltonian for a realistic (with charge noise)
Cooper pair box, one usually assumes that, irrespective of the exact 
microscopic nature, the charge impurities are charges
that tunnel between two positions. Each charge is described
as a quantum Two Level System (TLS) that is
controlled by the Hamiltonian  \cite{Anderson,Black-Halperin,Burin1995} :
\begin{equation} 
H_{TLS}=\epsilon \tau_z+ t \tau_x +H_{env} \;.
\label{qq}
\end{equation} 
Here $\epsilon$ is the energy difference between the two tunneling minima, 
$t$ is the tunneling amplitude and $\tau_{z,x}$ are the 
Pauli matrices acting on the impurity states.  Each TLS is subject to 
dissipation due to its own bath with Hamiltonian $H_{env}$.
In this picture, the interaction between several charged impurities and 
the charge states of the Cooper pair box can be written as:
\begin{equation} 
H_{SB}=\sum_{i} v_i \tau_z^i \sigma_z \;.
\label{int}
\end{equation} 
By adding Eqs. (\ref{Cpb}), (\ref{qq}) and (\ref{int}), we finally obtain  
the Hamiltonian of  a non-ideal (with charge noise) Josephson charge qubit.

Fortunately, experiments performed on Cooper pair boxes indicate that 
this rather complicated model of the charge noise can be simplified. 
In fact, it has been observed that charge impurities that are responsible 
for the low frequency noise behave 
essentially {\em classical}, i.e. they become equivalent to classical 
fluctuators each characterized by a switching rate 
$\gamma_i$ and the coupling to the qubit $v_i$. 
In this approximation the interacting Hamiltonian given in Eq. 
(\ref{int}) simplifies to the following qubit-fluctuators Hamiltonian:
\begin{equation}
H_{SB}^{cl}=\sum_{i} v_i \xi_i(t) \sigma_z \;
\label{intflu}
\end{equation}
and the quantum TLS dynamics given in Eq. (\ref{qq}) is substituted
by the classical noise $\xi_i(t)$, representing a 
symmetric Random Telegraph Process (RTP)  
switching between values $\pm 1 $ with rate $\gamma_i$ \cite{Joaknote}. 
If we assume a distribution $P(\gamma) \propto 1/\gamma$ at $\gamma \le \gamma_M$ for the switching rates, we find for the total noise 
${\chi (t)=\sum_i v_i \xi_i(t)}$ acting on the qubit a $1/f$ spectrum
$S(|\omega|) = A/|\omega|$, $\omega \le \gamma_M$, with coefficient  
$A$ proportional to the number of fluctuators per noise 
decade $n_d$ and their typical coupling to the qubit 
$\langle v \rangle^2$ \cite{Dutta1981,Weissman1988}. Note that we assume the 
switchings to be symmetric, that is, we assume the switching rate from the 
upper to the lower and from the lower to the upper TLS state to be the 
same. This is valid if the temperature is larger than the TSL level spacing
so that there is no population difference between the two states. This 
restriction can be relaxed \cite{Paladino2002} but for simplicity we 
shall work in the symmetric (high temperature) limit. 

Therefore, in this classical picture, 
the realistic Hamiltonian of a Cooper pair box afflicted by $1/f$ 
charge noise,  is given by the sum of Eq. (\ref{Cpb}) and 
Eq. (\ref{intflu}). In the qubit eigenstates, we find that:
\begin{equation}
H=\frac{\Delta E}{2} \rho_z + \frac{\chi(t)}{2} \left [\cos \theta \rho_z - \sin \theta \rho_x \right ] \;,
\label{Hqubit}
\end{equation}
where ${\Delta E=\sqrt{\delta E_c^2+E_J^2}}$,
${\theta =\arctan(E_J/\delta E_c)}$ defines the qubit working point and
$\rho_{x,z}$ are the Pauli matrices in the rotated basis.

The state of the qubit given in Eq. (\ref{Hqubit}) can 
be interpreted as the state of a pseudospin in a static magnetic field 
$\vec{B}=(0,0,\Delta E)$ along $\hat{z}$ with noise 
both in the $x$ and $z$ directions. Manipulations of the Cooper 
pair box state can be then achieved by 
applying an electromagnetic field which rotates in the 
$\hat{x}-\hat{y}$ plane at 
frequency $\omega_{rf}$, at or near the pseudospin precession 
frequency $\Delta E$ \cite{Collin2004}. 
Given the Hamiltonian of the control radio-frequency (RF) field as follows:
\begin{equation}
H_{rf}(t)=\Omega(t) \left [\cos(\omega_{rf} t+\varphi) \rho_x + \sin (\omega_{rf} t + \varphi) \rho_y \right ] \;
\label{rfHam}
\end{equation}
where $\varphi$ is the phase of the RF field and $\Omega(t)$ is the amplitude,
the total Hamiltonian  
$H^{tot}$ of the Josephson charge qubit with charge noise and applied control field reads (in the rotating frame):
\begin{eqnarray}
&&H^{tot}= \frac{\Delta E-\omega_{rf}}{2} \rho_z + \Omega (t) \bigl [\cos \varphi \rho_x + \sin \varphi \rho_y \bigr ] \;\label{Saclay} \\
&+& \frac{\chi(t)}{2} \bigl [\rho_z \cos \theta + \bigl ( \rho_x \cos(\omega_{rf} t) - \rho_y \sin(\omega_{rf} t) \bigr ) \sin \theta \bigr ]\;
\notag
\end{eqnarray}
In superconducting qubits typical values for the energy splitting are ${\Delta E \simeq 5-10}$ Ghz while the amplitude of the applied 
RF field $\Omega_{rf}$ might extend up to 250 MHz.

Different pulses sequences can be then generated by modulating the amplitude 
$\Omega(t)$ in Eq. (\ref{Saclay}).  Specifically, usual dynamical 
decoupling sequences \cite{Viola1998} are realized first by setting: \[
\Omega(t)=\sum_{k=1}^{N} \Omega_{rf} \left [\vartheta (t-k \tau) - \vartheta (t-k \tau - \tau_P)\right ] ;\]
where $N$ is the number of pulses, $\tau_P$ is the length of each pulse, 
$\vartheta(x)$ denotes the Heaviside step function and $\tau$ is the time 
interval between each pulse. Then, by choosing different values for the 
constant ${{\cal A}  =2 \Omega_{rf} \tau_P}$ and the phase 
$\varphi$. For example, rotations by $\pi$ around the $x$ or $y$ axes, 
$\pi_x$ and $\pi_y$ pulses, are achieved by fixing  
$\cal{A}=\pi$ and by setting respectively $\varphi=0$ and $\varphi=\pi/2$. 
Notice that, the so called bang-bang control, characterized by pulses that 
are infinitely short and strong so that there is no qubit dynamics 
during the time of the pulse, is achieved in the limit 
$\tau_P\rightarrow0$ for fixed value of the constant $\cal{A}$.
In this paper we always consider only this limiting case. 

\section{Ideal dynamical decoupling: exact solutions}
\label{ideal}
In this section we solve analytically the dynamics of a 
Cooper pair box coupled to a single fluctuator and subject to {\em ideal} 
(i.e. without errors in the pulses) dynamical decoupling both at pure 
dephasing and at the optimal working point for the qubit. 

The analytical solutions are derived by following the reasoning
given in Ref.~\onlinecite{Bergli2006}. 
The starting point is the Hamiltonian given in Eq. (\ref{Hqubit}) that describes
a qubit in interaction with a single fluctuator. The charge noise is then
given by $\chi(t)=v \xi(t)$, where $v$ is the coupling strength of 
the fluctuator and $\xi(t)$ denotes a symmetric RTP characterized by the
switching rate $\gamma$. 

It may be useful to recall here that in general,
depending on the ratio between $v$ and $\gamma$, we can distinguish 
between two different types of fluctuators: those characterized 
by ${v < \gamma}$ (named weakly coupled or Gaussian fluctuators), 
and those with ${v > \gamma}$ (named strongly coupled fluctuators) 
that are responsible for the non-Gaussian features of the noise \cite{Lara}. 
The effect on the qubit dynamics of weak and strong fluctuators is 
dramatically different. In fact, while the weak fluctuators are fast and 
produce essentially the same effects on the qubit as a bath of harmonic 
oscillators, thereby leading to the exponential decay of the coherent qubit 
oscillations at long times, the strongly coupled fluctuators, being slow,  
are responsible for memory effects in the qubit dynamics, for a fast initial 
decoherence of the qubit signal and for the inhomogeneous broadening of 
the qubit spectrum.

Due to the time dependence of the noise generated by the fluctuator, the 
Cooper pair box behaves as a pseudospin in a {\em variable} 
magnetic field. Therefore, solving its dissipative dynamics is a 
nontrivial task. 
Fortunately, because of the {\em bistable} nature of the 
fluctuator, the problem can be simplified. In fact, at different times, 
depending on the state of the fluctuator, 
the qubit vector state precesses on the Bloch sphere
around one of the two effective {\em static} magnetic fields: 
${\vec{B}_{\pm}=(\pm v \sin \theta,0, \Delta E \pm v \cos \theta)}$.
As a result, a convenient strategy to study the decoherence of 
the qubit is following the evolution of the  state vector
on the Bloch sphere. In this picture, 
solving the dissipative dynamics of the qubit  reduce to 
calculating the probability $p({\bf x},t)$ that at time $t$ the  
state vector reaches a point ${{\bf x} = (x,y,z)}$ on the sphere.

\subsection{Dynamical decoupling at pure dephasing.}
\label{idealpd}
At pure dephasing, ${\theta =0}$, the analytic solutions for the
dephasing rate $\Gamma_2$ of a qubit interacting with a single fluctuator 
are well known both in the case of free induction decay 
\cite{Lara,Galperin2004,Bergli2006} and when sequences of $\pi$ pulses are 
applied to the qubit \cite{Faorodec,Falcibb}.
Here we derive these results by using a different approach that allows us (i)
to simplify greatly the previous analytical solutions, (ii) to elucidate better
the diagnostic capability of the dynamical decoupling and finally (iii) 
to suggest an easy experiment that might be performed in order 
to extract information on the microscopic nature of the fluctuator.

At pure dephasing 
the qubit acquires an extra random phase 
shift ${\phi(t)=v \int_0^t dt' \chi(t')}$ due to the interaction with 
the fluctuator. The dephasing rate 
$\Gamma_2$ is then given by the decay rate of the average value:
\begin{equation}
\langle e^{i \phi} \rangle = \int d \phi p(\phi,t) e^{i \phi} \;.
\label{deph}
\end{equation}
In order to evaluate it, we need to calculate the 
probability distribution $p(\phi,t)$.

Let us first introduce 
the small time steps ${t_S \equiv t/M}$ and discretize the time 
integral given in Eq. (\ref{deph}). 
The random phase shift can be written as 
${\phi(t)=v t_S \sum_{n=1}^M \chi_n}$, where
${\chi_n \equiv \chi(n t_S)}$ and  the integration over time can be  
thought of as random walk process, where at each time step the random 
walker moves a step $x_S=v t_S$ in a direction depending on 
the current position of the fluctuator. 

To first order in $t_S$, the probability that the fluctuator 
does not switch, i.e. that the next step is made in the same direction as 
the previous one, is given by ${\alpha=P_0(t_S)=1-\gamma t_S}$.
Similarly, ${\beta =P_1(t_S)=\gamma t_S}$ is the probability that 
one switch occurs, i.e. that the step is made in the opposite direction of 
the previous one. Higher numbers of switches are negligible to the 
first order. 

Let us denote by $m$ the number of steps from the origin (i.e. 
the position of the random walker is $x=m x_S$) and by $n$ the number 
of temporal steps (i.e. the dimensional time is $t_n=n t_S$). Let us calculate the discretized probability $P_n(m)$ that 
the random walker reaches the position $m$ at time step $n$. It is useful to introduce the probabilities 
${P^+_n(m)}$ and ${P^-_n(m)}$ for the random walker to reach respectively 
the position $m$ coming from the left and from the right. 
These probabilities satisfy the following master equations:
\begin{equation}
\begin{split}\label{masterdeph}
P^+_{n+1}(m)&=\alpha P^+_n(m-1)=\beta P^-_n(m-1)\;\\
P^-_{n+1}(m)&=\beta P^+_n(m+1)=\alpha P^-_n(m+1) \;
\end{split}
\end{equation}
and are such that ${P_n(m)=P^+_n(m)+P^-_n(m)}$. 

In the limit of infinitesimal steps $t_S$, these probabilities reduce to:
\begin{eqnarray}
P^+_n(m)&=&p(m x_S,n t_S)=p_+(\phi,t) \; \notag \\
P^-_n(m)&=&p(m x_S,n t_S)=p_-(\phi,t) \; \notag.
\end{eqnarray}
Accordingly, the master equations given in Eqs. (\ref{masterdeph}) can be written as:
\begin{equation}
\partial_t {\bf p}(\phi,t)=M_{deph} {\bf p}(\phi,t) \;
\label{eqp}
\end{equation}
where we have introduced the matrix:
\begin{equation}
M_{deph}=\left (\begin{array}{cc} -\gamma -v \partial_\phi & \gamma \\ \gamma & - \gamma + v \partial_\phi \end{array} \right ) \;
\label{meqp}
\end{equation} 
and we have defined: 
${{\bf p}(\phi,t)=\left (\begin{array}{ll} p^+(\phi,t) \\ p^- (\phi,t)\end{array} \right )}$.

The solutions of Eqs. (\ref{eqp}) can be easily guessed
as
${{\bf p} = {\bf C} e^{i \kappa \phi-\omega t}}$ and we get the 
following dispersion equation:
\begin{equation} 
-i\omega{\bf C}=\tilde{M}_{deph}{\bf C}\;
\label{dis1}
\end{equation}
where the matrix
\begin{equation}
 \tilde{M}_{deph} = \left( 
    \begin{array}{cc}
      -\gamma-iv \kappa& \gamma\\
      \gamma&-\gamma+iv \kappa
    \end{array} \right) \;
\end{equation}
has the following eigenvalues:
\begin{equation} 
  \omega_\pm = -i\gamma\pm i \gamma\sqrt{1-\frac{v^2}{\gamma^2} \kappa^2}
    = -i\gamma\pm i\gamma\mu_\kappa
\end{equation}
and corresponding normalized eigenvectors:
\begin{equation}
  {\bf v}_\pm = \frac{1}{\sqrt{1+(-i\frac{v}{\gamma}\kappa \pm\mu_\kappa)^2}}
  \left(\begin{array}{c} -i\frac{v}{\gamma}\kappa\pm\mu_\kappa\\1\end{array}\right)
\end{equation}
It follows that the general solution to Eq. (\ref{eqp}) reads:
\begin{eqnarray}
  {\bf p}(\phi,t)& =&\int_{-\infty}^{\infty} \frac{d \kappa}{2 \pi} \left [
A_+ {\bf v}_+e^{-i\omega_+t}+A_-{\bf v}_-e^{-i\omega_-t} \right ] 
e^{i \kappa \phi} \, \notag \\ 
&=&\int_{-\infty}^{\infty} \frac{d \kappa}{2 \pi} {\bf \tilde{p}}(t) e^{i \kappa \phi} \; ,
\label{sol}
\end{eqnarray}
where the constants $A_\pm$ are determined by the initial conditions: 
\begin{equation}
  {\bf\tilde{p}}(0) = A_+{\bf v}_+  + A_-{\bf v}_-\;,
\label{is}
\end{equation}
or, by using a more compact notation, ${A_\pm =  {\bf\tilde{p}}(0)\cdot{\bf v}_\pm}$.

By introducing the matrices:
\begin{equation}
 U=\left(\begin{array}{cc}e^{-i\omega_+t}&0 \\0&e^{-i\omega_-t}
\end{array}\right) ~~~ \text{and}~~~
   V = ({\bf v}_+|{\bf v}_-)
\end{equation}
we can conveniently write Eq. (\ref{sol}) as follows:
\begin{equation}
{\bf p}(\phi,t) =\int_{-\infty}^{\infty} \frac{d \kappa}{2 \pi} \left [ \Lambda(t) {\bf\tilde{p}}(0) \right ]
e^{i \kappa \phi}  \;
\label{sol1}
\end{equation}
where ${\Lambda(t)=VUV^T}$. 
Then, by substituting Eq. (\ref{sol1}) into Eq. (\ref{deph}), we finally find that the average phase shift reads:
\begin{eqnarray}
\langle e^{i\phi} \rangle &=& \int_{-\infty}^{\infty} d \kappa \int \frac{d \phi}{2 \pi} \left [ \Lambda (t) {\bf\tilde{p}}(0) \right ]
e^{i (\kappa+1) \phi}\; \notag \\
&=&  \int_{-\infty}^{\infty} d \kappa \left [ \Lambda (t){\bf\tilde{p}}(0) \right ] \delta(\kappa+1) \nonumber \\
&=& \left [ \Lambda (t){\bf\tilde{p}}(0) \right ]_{\kappa=-1}\;.
\label{deph1}
\end{eqnarray} 
In order to evaluate the dephasing rate $\Gamma_2$, it is convenient to expand
the vector ${\bf\tilde{p}}(0)$ in terms of the  eigenvectors of $\Lambda (t)$.
It is then a matter of simple algebra to find that the resulting decay of the average phase shift given in Eq. (\ref{deph1}) is indeed given by the
sum of two exponentials with coefficients
$\Gamma_2^\pm= -\ln|\lambda_\pm|/t$, where
\begin{equation}
\lambda_\pm=e^{-\gamma t \pm \gamma\mu t} \;
\end{equation}
are the eigenvalues of the matrix $\Lambda(t)$ and we define 
${\mu \equiv \mu_{-1}=\sqrt{1-\frac{v^2}{\gamma^2}}}$.

The long time decay is given by the component that decays slower, i.e. the
one that corresponds to the larger eigenvalue
\begin{equation}
\Gamma_2=-\ln|\max(\lambda_+,\lambda_-)|/t = \gamma \left (1-\Re \mu \right ) \;
\end{equation}
This result  agrees with the one given in 
Refs. \cite{Lara,Galperin2004,Bergli2006}. 

A similar approach can be used in order to evaluate the dephasing rate
$\Gamma_2^{\text{dd}}$ of the qubit that is subject to dynamical decoupling.
In particular, let us assume that we apply to the qubit a sequence of  
$\pi_x$ pulses.  It is easy to realize that the effect of each $\pi_x$ pulse 
is simply to interchange the constants $A_+$ and $A_-$.  
As a result, when a sequence of $N$ 
$\pi_x$ pulses is applied, the probability given in Eq. (\ref{sol1}) 
reduces to:
\begin{equation}
{\bf p}(\phi,t) =\int_{-\infty}^{\infty} \frac{d \kappa}{2 \pi} \left [ 
\Lambda^{\text{dd}}(t) {\bf\tilde{p}}(0) \right ]
e^{i \kappa \phi}  \;
\label{sol1bb}
\end{equation}
where:
\begin{equation}
\Lambda^{\text{dd}}(t)= 
  \sigma_x \Lambda (\tau) \cdots \sigma_x \Lambda (\tau) \;
\end{equation}
and $t=N\tau$ is the total time. 

Once this probability is known, it is straightforward to evaluate first 
the average phase shift acquired by the qubit:
\begin{eqnarray}
\langle e^{i\phi} \rangle &=& \int_{-\infty}^{\infty} d \kappa \int \frac{d \phi}{2 \pi} \left [ \Lambda^{\text{dd}}(t) {\bf\tilde{p}}(0) \right ]
e^{i (\kappa+1) \phi}\;\label{deph2} \\
&=&  \int_{-\infty}^{\infty} d \kappa \left [ \Lambda^{\text{dd}} (t){\bf\tilde{p}}(0) \right ] \delta(\kappa+1) 
 = \left [ \Lambda^{\text{dd}} (t){\bf\tilde{p}}(0) \right ]_{\kappa=-1}\;.
\notag
\end{eqnarray} 
and then the corresponding dephasing rate:
\begin{equation}
\Gamma_2^{\text{dd}}=-\ln|\max(\tilde{\lambda}_+^N,\tilde{\lambda}_-^N)|/t \;.
\label{depbb}
\end{equation}
Here ${\tilde{\lambda}_\pm}$ are the eigenvalues of ${\sigma_x \Lambda(\tau)}$.
Depending on the nature of the fluctuator (i.e. weak or strong), 
the eigenvalues ${\tilde{\lambda}_\pm}$ can be  written as a function both of the fluctuator and  the control parameters.

Let us first consider that the fluctuator coupled to the qubit is weak, i.e. 
${v<\gamma}$. We find that:
\begin{equation}
  \tilde{\lambda}_\pm = \frac{e^{-\gamma\tau}}{\mu}\left[\sinh\gamma\mu\tau
      \pm\sqrt{\cosh^2\gamma\mu\tau-\frac{v^2}{\gamma^2}}\right] \;
\label{autw}
\end{equation}
Since the eigenvalue $\tilde{\lambda}_+$ is 
the maximum one for each choice of the parameters, we find that 
the dephasing rate is given by:
\begin{equation}
\Gamma_2^{\text{dd}} = -\ln|\tilde{\lambda}_+^N|/t=-\ln|\tilde{\lambda}_+|/\tau \;
\label{bbw}
\end{equation}
In particular, to the lowest order in $v/\gamma$ we find that:
\begin{equation}
\Gamma_2^{\text{dd}} = \frac{v^2}{2\gamma}\left[1-\frac{1}{\gamma\tau}\tanh\gamma\tau
\right] \;.
\label{bbwl}
\end{equation}
From this simple analytical result, it is evident that in order 
to achieve an efficient suppression of the noise generated by a weak 
fluctuator we need to choose the time interval between two pulses 
$\tau\lesssim1/\gamma$. Namely, the efficiency of the dynamical decoupling
is related to the switching rate $\gamma$ of the weak fluctuator.

\begin{figure}[h]
\begin{center}
\psfrag{x}{$\tau$}
\psfrag{y}{$\Gamma_2^{\text{dd}}/\Gamma_2$}
\includegraphics[width=7cm]{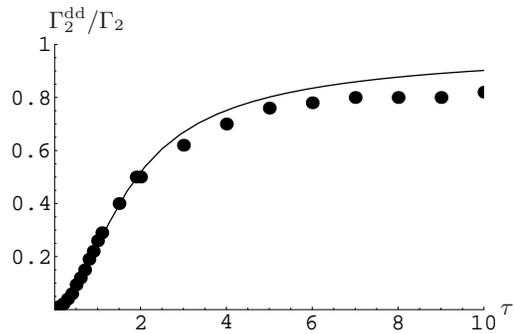}
\end{center}
\caption{
Dephasing rate of a qubit interacting with a {\em weak} fluctuator as 
a function of the time interval $\tau$ of the bang bang pulses. The
analytical solution given in Eq. (\ref{bbw}) is represented by the solid line, 
while the black dots indicate the numerical results. 
The values for the fluctuator parameters are respectively 
$v=0.1$ and $\gamma=1$ (in unit of $\Delta E$). 
The dephasing rate $\Gamma_2=v^2/2\gamma$ is given in Eq. (\ref{bbwl}).\label{f1}}
\end{figure}

In Fig. \ref{f1} we plot $\Gamma_2^{\text{dd}} (\tau)$ given in Eq. (\ref{bbwl}) and 
we compare our analytic solution with the result of a numerical 
simulation where the time evolution of the qubit was followed for a given 
realization of the fluctuator RTP and averages taken over many such 
realizations. Numerical and analytical results are in good agreement.
Moreover, as one should expect, by reducing the time interval between the 
pulses, the dephasing rate $\Gamma_2^{\text{dd}}$ vanishes.  

Let us now consider the case of a strongly coupled fluctuator.
For $v> \gamma$, the constant $\mu$ becomes complex 
($\mu=i\tilde{\mu}$) and we find that the eigenvalues read now:
\begin{equation}
   \tilde{\lambda}_\pm =
  \frac{e^{-\gamma\tau}}{\tilde{\mu}}\left[\sin\gamma\tilde{\mu}\tau
      \pm\sqrt{\frac{v^2}{\gamma^2}-\cos^2\gamma\tilde{\mu}\tau}\right] \;
\end{equation}
Differently from the expressions given in Eqs. (\ref{autw}), here the
two eigenvalues ${\lambda_{\pm}}$  are oscillating in the time 
interval $\tau$ between the pulses. It follows that there are two relevant 
dephasing times now:
\begin{equation}
\Gamma_2^{\text{dd}\pm} = -\ln|\tilde{\lambda}_\pm^N|/t.
\label{bbs}
\end{equation} 
To the lowest order in $\gamma/v$, we find that:
\begin{equation}
 \Gamma_2^{\text{dd}\pm}= -\ln|\lambda_\pm|/\tau= 
\gamma\left[1\pm\frac{1}{v\tau}\sin v\tau
\right] \;
\label{bbsl}
\end{equation}
Notice that, by looking at Eq. (\ref{bbsl}), it is quite
straightforward to realize that 
for the case of strong fluctuator an efficient 
suppression of the noise is achieved by setting $\tau\lesssim 1/v$. Therefore,
the efficiency of the dynamical decoupling is related to the 
coupling strength $v$ of the strong fluctuator.

\begin{figure}[h]
\begin{center}
\psfrag{x}{$\tau$}
\psfrag{y}{$\Gamma_2^{\text{dd}}/\Gamma_2$}
\psfrag{a}{$\Gamma_2^{\text{dd}-}$}
\psfrag{b}{$\Gamma_2^{\text{dd}+}$}
\includegraphics[width=7cm]{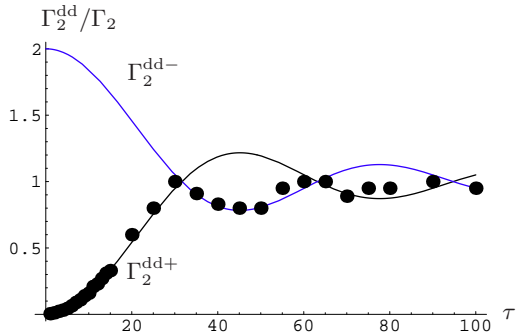}
\end{center}
\caption{
Dephasing rates of a qubit interacting with a {\em strong} 
fluctuator as a function of the time interval $\tau$ of the bang bang pulses. 
Solid lines represent the analytical solutions given in Eqs. (\ref{bbs}) while black dots 
are the numerical results. The values for the fluctuator 
parameters are respectively $v=0.1$ and $\gamma=0.01$ (in units of
$\Delta E$). From Eq. (\ref{bbsl}) we have $\Gamma_2=\gamma$.\label{f2}}
\end{figure}

In Fig.~\ref{f2} we plot the dephasing rates 
${\Gamma_2^{\text{dd}\pm}(\tau)}$ given in Eq.~(\ref{bbs}) and we compare 
them with numerics.
As evidenced by Fig.~\ref{f2}, the results of the numerical simulations 
clearly show that, depending on the 
time interval $\tau$ between the pulses, the dephasing rate ${\Gamma_2^{\text{dd}}}$ of the qubit is dominated differently by the two dephasing rates 
${\Gamma_2^{\text{dd}\pm}}$. 

In order to understand this behavior it is useful to write
the average phase shift given in Eq. (\ref{deph2}) as follows:
\begin{equation}
\langle e^{i \phi} \rangle = w_+ \tilde{\lambda}_+^N  +  
w_- \tilde{\lambda}_-^N \;,
\label{pesi}
\end{equation}
where we have introduced the weights $w_\pm$. These can be calculated 
as follows. First, we write the initial state 
$ {\bf\tilde{p}}(0)$, given in Eq. (\ref{is}), as 
a linear superposition of the eigenvectors ${\bf \tilde{u}}_\pm$ 
of the matrix $\sigma_x \Lambda (\tau)$: 
\[ {\bf\tilde{p}}(0) = \sum_{i=\pm}B_i{\bf \tilde{u}}_i .\] 
Then, we solve the following
linear system of equations in order to find the coefficients $B_i$:
\begin{equation}
{\bf\tilde{p}}(0)\cdot{\bf \tilde{u}}_j=\sum_i B_i{\bf \tilde{u}}_i\cdot{\bf \tilde{u}}_j.
\end{equation}
For example, in the numerical simulations shown in Fig.~\ref{f2}, 
the fluctuator was initially assumed at equilibrium, i.e.
${\bf\tilde{p}}(0) =(1,1)/2$. By following the above procedure, we find that
the weights read: 
\begin{equation}
 w_\pm=\frac{1}{2}\pm\frac{\mu\cosh\gamma\mu\tau}
   {2\sqrt{\cosh^2\gamma\mu\tau-\frac{v^2}{\gamma^2}}}.
\label{weights}
\end{equation}
In particular, to the lowest order in $\gamma/v$, we obtain that 
${w_\pm (\tau) = \frac{1}{2}(1\pm\cos v\tau)}$ and we can immediately  
explain the oscillating behavior displayed in Fig.~\ref{f2}.

An important consideration is relevant at this point.
By comparing  Fig.~\ref{f1} and Fig.~\ref{f2} it is immediate to realize
that the two plots are {\em qualitatively} different. 
One can then exploit this difference in order to 
distinguish between a weak and a strong fluctuator 
coupled to the qubit. For example, in an experiment, once a single 
flucutator has been singled out in a device, one could measure  
systematically the dephasing time of the qubit after cycles of pulses 
characterized by different times $\tau$ and observe if there is 
an oscillatory or not behavior in the data. 
The efficiency of the dynamical decoupling will then provide further 
information respectively on the coupling strength of the fluctuator $v$ 
or its switching rate $\gamma$.

However, in order to characterize completely the fluctuator both the 
two parameters $v$ and $\gamma$ must be known. 

\subsection{Dynamical decoupling at the optimal point.}
\label{idealop}

In this section we study the dissipative dynamics of a Cooper pair box 
tuned at the optimal point, i.e. ${\theta =\frac{\pi}{2}}$, that is subject 
to control sequences of $\pi$ pulses. 
By using a similar approach to the one discussed in the previous section, 
we find the analytic solution for the qubit decoherence rate. 

Contrary to the case of pure dephasing, at the optimal point 
the performance of the dynamical decoupling depends both on 
the characteristics of the charged fluctuators coupled to the qubit 
and on the qubit energy level splitting. 
In particular, quite worryingly, it has been noticed that the 
qubit decoherence can be even {\em accelerated} if the cycle time of the 
control is 
long with respect to the period of the free evolution of the qubit
\cite{Faorodec,Falcibb}. This acceleration is sometimes referred in the 
literature as the anti-Zeno effect \cite{Kofman}.

In the following we shall see indeed that this acceleration
appears only when sequences of $\pi_x$ pulses are used 
to flip the qubit state and  can be 
conveniently avoided by flipping the state with $\pi_y$ pulses.

To this aim, let us introduce the time evolution 
operator $\displaystyle{U(t)= T e^{-i \int_0^t H(t')d t'}}$, where
$H(t)$ is the time dependent Hamiltonian given in Eq.~(\ref{Hqubit}) and 
let us study the time evolution of the qubit when a cycle of 
$\pi_x$ and $\pi_y$ pulses is applied to it.
Let $\tau$ be the time interval between the pulses and let us consider 
first the case of pure dephasing. 

Since the time evolution operator $U(t)$ depends only on 
the qubit operator $\sigma_z$, at pure dephasing we see that the 
noise coming from fluctuators that are static during the time interval 
$\tau$ can be completely eliminated either by applying a cycle of 
$\pi_x$ or $\pi_y$ pulses. In fact, mathematically we obtain that:
\begin{equation}
\sigma_x U(\tau) \sigma_x U(\tau) = \sigma_y U(\tau) \sigma_y U(\tau) = \openone\;
\end{equation}
where $\openone$ is the identity.

On the contrary, at the optimal point, the evolution operator 
$U(t)$ depends both on the qubit operator $\sigma_z$ 
(because of the interaction with the fluctuators) and $\sigma_x$ 
(because of the Josephson coupling energy $E_J$). Mathematically
we find that:
\begin{equation}
\sigma_x  U(\tau) \sigma_x U(\tau) \neq \openone ~~~\text{and}~~~ \sigma_y U(\tau) \sigma_y U(\tau) = \openone
\end{equation}
Namely, only  by flipping the qubit state around the 
$y$-axis we can completely suppress the noise 
due to fluctuators that are static in the time interval $\tau$. 
Therefore, at the optimal point, the 
performance of the dynamical decoupling changes remarkably depending whether
sequences of $\pi_x$ or $\pi_y$ pulses are used to decouple the qubit.

\begin{figure}[h]
\psfrag{r}{$\tau$}
\psfrag{s}{$\Gamma_2^{\text{dd}}/\Gamma_2$}
\psfrag{t}{$\tau$}
\psfrag{w}{$\Gamma_2^{\text{dd}}/\Gamma_2$}
\psfrag{a}{$\pi_y$- Pulses}
\psfrag{b}{$\pi_x$- Pulses}
\includegraphics[width=7cm]{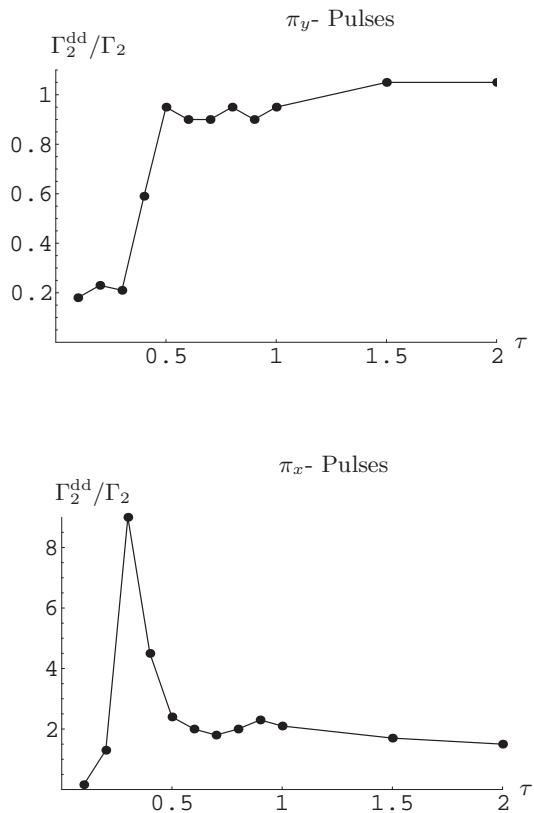}
\caption{Dissipative dynamics of the dynamical decoupling applied to 
a Cooper pair box coupled to a {\em weak} fluctuator at the optimal point
($E_J=10$, $v=0.1$, $\gamma=1$). From Eq. (\ref{w11}) we have 
$\Gamma_2=\gamma v^2/E_J^2$. The lines are added to guide the eye and 
do not represent any analytical result  
\label{f8w} }
\end{figure}

\begin{figure}[h]
\psfrag{x}{$\tau$}
\psfrag{y}{$\Gamma_2^{\text{dd}}/\Gamma_2$}
\psfrag{u}{$\tau$}
\psfrag{v}{$\Gamma_2^{\text{dd}}/\Gamma_2$}
\psfrag{a}{$\pi_y$- Pulses}
\psfrag{b}{$\pi_x$- Pulses}
\includegraphics[width=7cm]{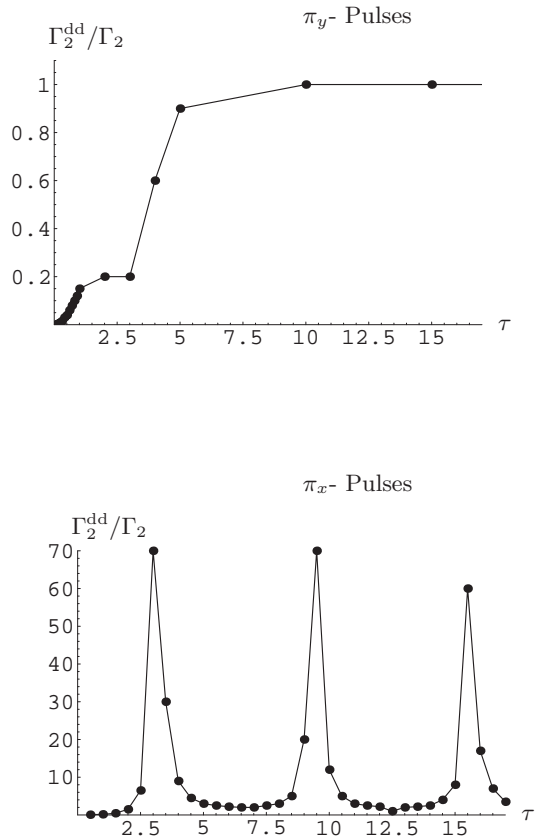}
\caption{Dissipative dynamics of the dynamical decoupling applied to 
a Cooper pair box coupled to a {\em strong} fluctuator at the optimal point
($E_J=1, v=0.1, \gamma=0.01$).
From Eq. (\ref{sss}) we have $\Gamma_2=\gamma v^2/E_J^2$. 
\label{f8s}}
\end{figure}

These considerations are clearly illustrated in Fig.~\ref{f8w} and 
Fig.~\ref{f8s}. In Fig.~\ref{f8w} we study numerically 
the dissipative dynamics of a 
Cooper pair box that is coupled to a weak fluctuator. The qubit 
dephasing rate $\Gamma_2^{\text{dd}}$ is displayed as a function of the 
time interval $\tau$ between cycles of $\pi_y$ pulses (upper Panel) and 
$\pi_x$ pulses (lower Panel).
Notice that only when $\pi_y$ pulses are applied to the qubit, the dephasing 
rate $\Gamma_2^{\text{dd}}$ is monotonously reduced as the flipping rate 
is increased. On the contrary, for sequences of $\pi_x$ pulses, there are 
values of $\tau$ where the dephasing rate $\Gamma_2^{\text{dd}}$ increases 
remarkably as compared to the rate when there are no control-flips 
(i.e. anti-Zeno effect \cite{Faorodec,Falcibb}). 

A similar analysis is performed in Fig.~\ref{f8s} where the dissipative 
dynamics of the qubit coupled to a strong fluctuator is illustrated.
Again, we find that the dephasing rate $\Gamma_2^{\text{dd}}$ displays both
the anti-Zeno effect and rapid oscillations  
when sequences of $\pi_x$ pulses are applied to the qubit (lower Panel).
However, both the the acceleration and the oscillations disappear when control 
sequences of $\pi_y$ pulses are instead used (upper Panel). 

Note that when control sequences of $\pi_x$ pulses are used 
both in  Fig.~\ref{f8w} and 
Fig.~\ref{f8s} the efficiency of the dynamical decoupling is related to the 
qubit energy scale $E_J$. 
In this case, in fact we need to choose time interval 
${\tau \lesssim 1/E_J}$ in order to obtain an efficient 
suppression of the noise due to the fluctuator. 
For sequences of $\pi_y$ pulses the 
efficiency of the dynamical decoupling is determined either by the 
switching rate ${\tau \lesssim 1/\gamma}$ (if $\gamma<E_J$) 
or energy scale ${\tau \lesssim 1/E_J}$ (if $\gamma>E_J$) for the 
weak fluctuator and by 
${\tau \lesssim 1/\gamma}$) for the strong fluctuator. This will be 
discussed after we have derived the analytic expressions below.

As a final remark, let us notice that the different performances of sequences of 
$\pi_x$ and $\pi_y$ pulses can be easily explained in the context of effective Hamiltonian 
theory \cite{Chuang2005}, by stating that while $\pi_x$-pulses provides 
decoupling only to the first order in the Magnus expansion of the 
dephasing rate, $\pi_y$-pulses gives exact decoupling to all orders. 

Having clarified this point, let us now proceed to derive the analytical 
solution for the dissipative dynamics of the qubit at optimal point.
In view of the above considerations, in the following we shall 
only consider control sequences of $\pi_y$ pulses.

The analytical solution is derived by following the same reasoning discussed 
in the previous section. 
Here, the main difference is that the two effective static magnetic 
fields around which the state of the Cooper pair box precesses 
read now: ${\bf B}_\pm=(\pm v,0,E_J)$. As a consequence, at the optimal point,
the qubit does not acquire only a simple random phase due to the 
interaction with the charged fluctuator, 
but its state will spread over the entire Bloch sphere. 

In order to calculate the probability $p({\bf x},t)$ that the qubit
vector-state reaches the point ${\bf x}=(x,y,z)$ on the Bloch sphere at a 
certain time $t$,
first we split the probability ${p({\bf x},t)=p_+({\bf x},t)+p_-({\bf x},t)}$ 
and then we derive the master equations for $p_+({\bf x},t)$ and 
$p_-({\bf x},t)$:
\begin{equation}
\begin{split}\label{master}
 p_+({\bf x},t+t_S)
  & =  \alpha p_+(U_+^{-1}{\bf x},t)+\beta p_-(U_+^{-1}{\bf x},t) \\
 p_-({\bf x},t+t_S) 
  & =  \beta p_+(U_-^{-1}{\bf x},t)+\alpha p_-(U_-^{-1}{\bf x},t)
\end{split}
\end{equation}
where ${\alpha = 1-\gamma t_S}$ is the probability for the fluctuator 
to remain in the same state at the time $t_S$ while 
${\beta=\gamma t_S}$ denotes the probability that it switches. 
Notice that in Eqs. (\ref{master}) we have introduced the operator
$\displaystyle{U_\pm = e^{t_S {\bf B}_\pm \cdot {\bf R}}}$, where 
${{\bf R}=(R_x,R_y,R_z)}$ and the rotation matrices are defined as follows:
\[
  R_x = \left(\begin{array}{ccc}0&0&0\\0&0&-1\\0&1&0\end{array}\right)\\
  R_y = \left(\begin{array}{ccc}0&0&1\\0&0&0\\-1&0&0\end{array}\right)\\
  R_z = \left(\begin{array}{ccc}0&-1&0\\1&0&0\\0&0&0\end{array}\right)\\
\]
By performing the continuous limit $t_S \rightarrow 0$, from Eqs. 
(\ref{master}) we obtain directly the equations governing the time evolution of 
the Cooper pair box {\em before} the first $\pi_y$ pulse is applied:
\begin{equation}
\begin{split}\label{e1}
 \partial_t p_+ &= \gamma p_- -({\bf B}_+\cdot {\bf R}{\bf x})\cdot\nabla_{\bf x}p_+
     -\gamma p_+\\
 \partial_t p_- &= \gamma p_+-({\bf B}_-\cdot {\bf R}{\bf x})\cdot\nabla_{\bf x}p_-
     -\gamma p_-
\end{split}
\end{equation}
In this picture to apply a $\pi_y$ pulse to the qubit simply 
means to reverse the effective magnetic field. As a result, 
given Eqs.~(\ref{e1}), it is immediate to write 
the equations describing the time evolution of the state of the Cooper 
pair box {\em after} the application of the first $\pi_y$ pulse:
\begin{equation}
\begin{split}\label{e2}
 \partial_t p_+ &= \gamma p_- +({\bf B}_+\cdot {\bf R}{\bf x})\cdot\nabla_{\bf x}p_+
     -\gamma p_+\\
 \partial_t p_- &= \gamma p_+ +({\bf B}_-\cdot {\bf R}{\bf x})\cdot\nabla_{\bf x}p_-
     -\gamma p_-
\end{split}
\end{equation}
Clearly, once the second $\pi_y$ pulse is applied, 
the dissipative dynamics of the box is again described by the set 
of equations given in Eqs. (\ref{e1}) and so on for all the control cycles. 
Still we need to understand 
how to match the solutions just before and after the application of each
$\pi_y$ pulse.  

Let ${p_\pm({\bf x},\tau \to 0^-)}$ be
the solutions of Eqs.~(\ref{e1}) just before the 
first pulse, while ${p_\pm({\bf x},\tau \to 0^+)}$ 
are the solutions just after the pulse.  Notice that the 
latter are indeed initial conditions for Eqs. (\ref{e2}).
It is then easy to realize that a $\pi$ rotation 
around the $y$-axis leads to the following matching conditions:
\begin{equation}\label{yr}
\begin{split}
 p_+({\bf x},\tau \to 0^+) &= p_+({\bf x},\tau \to 0^-)\\
 p_-({\bf x},\tau \to 0^+) &= p_-({\bf x},\tau \to 0^-)
\end{split}
\end{equation}
Therefore, we can write explicitly Eqs.~(\ref{e2}) as follows: 
\begin{equation} 
\begin{split}\label{e3}
  \partial_t p_+ &= \left [E_J (y \partial_x  -x \partial_y ) + v  (z \partial_y   - y \partial_z ) \right ]p_+ - \gamma \delta p \\
  \partial_t p_- &=\left [ E_J (y \partial_x - 
x \partial_y) - v (z \partial_y - y \partial_z ) \right ] p_- + 
\gamma \delta p  
\end{split}
\end{equation}
where ${\delta p = p_+ - p_-}$.

In order to solve them, it is useful to define the following 
expectation values:
\begin{equation}
k_\pm = \int d^3 k p_\pm( {\bf x}, t) k \qquad k=x,y,z\;. \label{medie} 
\end{equation}
We can then write Eqs.~(\ref{e3}) in terms of Eqs.~(\ref{medie}) and we
obtain the following set of equations 
\begin{equation}
\begin{split}\label{s}
 \partial_t x_+ &= -\gamma \left (x_+ - x_- \right )  - E_J y_+\\
 \partial_t y_+  &= -\gamma \left ( y_+ - y_- \right ) + E_J x_+ -vz_+\\
 \partial_t z_+ &= -\gamma (z_+ - z_-) + v y_+\\
 \partial_t x_- &= \gamma (x_+ - x_-) - E_J y_-\\
 \partial_t y_- &= \gamma (y_+ - y_-) + E_J x_- +vz_-\\
 \partial_t z_- &= \gamma (z_+ - z_-) - v y_-\\
\end{split}
\end{equation}
having constant coefficients. Moreover, by introducing the variables 
${X_\pm \equiv x_+ \pm x_-}$, ${Y_\pm \equiv y_+ \pm y_-}$ and ${Z_\pm \equiv z_+ \pm z_-}$, 
Eqs.~(\ref{s}) can be further reduced into the following two independent sets of equations:
\begin{equation}
\begin{split}\label{s1}
 \partial_t X_+ &=  -E_J Y_+\\
 \partial_t Y_+ &= E_J X_+ -vZ_-\\
 \partial_t Z_- &=  -2\gamma Z_- + v Y_+\\
\end{split}
\end{equation}
and
\begin{equation}
\begin{split}
 \partial_t X_- &= -2\gamma X_- - E_J Y_-\\
 \partial_t Y_- &= -2\gamma Y_- + E_J X_- -vZ_+\\
 \partial_t Z_+ &=  v Y_- \\
\end{split}
\end{equation}
Finally, by solving the system of Eqs.~(\ref{s1}), we can estimate the 
dephasing rate $\Gamma_2$. To this aim, it is sufficient to prepare the 
system in an  equal superposition of eigenstates, i.e. with $Z=0$, 
and then to study the decay of the $x$ and $y$ components of the Bloch vector. 
Namely, we need to derive the decaying behavior of the variables $X_+$ and $Y_+$. 

It is convenient to write the system of Eqs.~(\ref{s1}) in matrix form:
\begin{equation}
  \partial_t\left(\begin{array}{c}X_+\\Y_+\\Z_-\end{array}\right)
 = M_1\left(\begin{array}{c}X_+\\Y_+\\Z_-\end{array}\right)
\end{equation}
where the matrix $M_1$ reads:
\begin{equation}
  M_1 = \left(\begin{array}{ccc} 0&-E_J&0\\E_J&0&-v\\0&v&-2\gamma\end{array}\right) \;
\end{equation}
After a $\pi_y$ pulse is applied to the qubit, the matrix $M_1$ is replaced by
the following matrix:
\begin{equation}
  M_2 = \left(\begin{array}{ccc} 0 &  E_J & 0\\-E_J&0&v\\0&-v&-2\gamma
\end{array}\right)
\end{equation}
By resorting to Eqs. (\ref{yr}), it is straightforward to see 
that the evolution of the qubit state after one full control cycle 
can be described by the operator: 
\begin{equation}
 T = \Pi_y e^{M_2\tau}\Pi_y e^{M_1\tau} \;\label{ev}
\end{equation}
where we introduced the matrix ${\Pi_y= \text{diag}(1,1,1).}$

Eq.~(\ref{ev}) can be further simplified to:
\begin{equation}
T=\left (\Pi_y L e^{M_1 \tau} \right )^2 \;
\label{ev2}
\end{equation}
where we defined the matrix ${L=\text{diag}(1,-1,1)}$. Finally we find that, 
after an echo cycle:
\begin{equation}
X_+ = \sum_{i=1}^{3} w_i e^{- \Gamma_2^{\text{dd}(i)} t} \qquad i=1,2,3\; \label{solu}
\end{equation}
where 
\begin{equation}
\Gamma_2^{\text{dd}(i)} = -\ln|\mu_i|/\tau \;
\label{optsol}
\end{equation} 
and $\mu_i$ are the eigenvalues of the matrix $T$. It follows that 
the dissipative dynamics of the qubit tuned at the optimal point is 
controlled by three different dephasing rates. 

Let us now discuss the decay rates in the limiting cases of 
weak ($v\ll\gamma$) and strong ($v\gg\gamma$) fluctuator. In both cases, it 
can be proved that in Eq. (\ref{solu}) the weight 
$w_3=0$. As a result, only the two dephasing rates $\Gamma_2^{\text{dd}(1)}$ 
and $\Gamma_2^{\text{dd}(2)}$ are relevant for the qubit dissipative dynamics.

Let us first evaluate them for the case of a {\em weak} fluctuator. 
After some calculations we find that:
\begin{widetext}
\begin{equation}\label{y1}
  \Gamma_2^{\text{dd}(1)} = \frac{\gamma v^2}{E_J^2+4\gamma^2}
   \left[1-\frac{E_J^2-4\gamma^2}{E_J^2+4\gamma^2}
         \frac{\sin \left (E_J \tau \right )}{E_J \tau} - \frac{8\gamma^2}{E_J^2+4\gamma^2}
  \cos^2 \left (\frac{E_J \tau}{2} \right )\frac{\tanh \left (\gamma \tau \right )}{\gamma \tau}\right]
\end{equation}
\begin{equation}\label{y2}
  \Gamma_2^{\text{dd}(2)} = \frac{\gamma v^2}{E_J^2+4\gamma^2}
   \left[1+\frac{E_J^2-4\gamma^2}{E_J^2+4\gamma^2}
         \frac{\sin \left ( E_J \tau \right )}{E_J \tau} - \frac{8\gamma^2}{E_J^2+4\gamma^2}
  \sin^2 \left (\frac{E_J \tau}{2} \right )\frac{\coth \left (\gamma \tau \right )}{\gamma \tau}\right]
\end{equation}
\end{widetext}
Both from Eq. (\ref{y1}) and Eq. (\ref{y2}) it is evident that, contrary to 
the case of pure dephasing, the energy level splitting of the qubit $E_J$ 
plays now a major role in the dissipative dynamics of the qubit.

In order to understand the efficiency  of the dynamical 
decoupling, it is useful to discuss the dephasing rates in two 
limiting case.

(i) {\em weak} fluctuator with ${\gamma \lesssim E_J}$.\\
In this case Eq. (\ref{y1}) and Eq. (\ref{y2}) reduce to:
\begin{eqnarray}
\Gamma_2^{\text{dd}(1)} &\approx& 
  \frac{\gamma v^2}{E_J^2}\left[1-\frac{\sin \left (E_J \tau \right )}{E_J \tau}\right] \label{w11} \\ 
\Gamma_2^{\text{dd}(2)} &\approx&
  \frac{\gamma v^2}{E_J^2}\left[1+\frac{\sin\left (E_J \tau \right )}{E_J \tau}
    -\frac{8\gamma^2}{E_J^2} \sin^2 \left ( \frac{E_J \tau}{2} \right ) 
\frac{\coth \left (\gamma\tau \right )}{\gamma \tau} \right] \notag
\end{eqnarray}
From Eq. (\ref{w11}) it is evident that in order to achieve an efficient 
suppression of the noise induced by the fluctuator we need to apply sequences 
of $\pi_y$ pulses with time interval between the pulses $\tau$ such 
that ${\tau \le \pi/E_J}$. Once this constrain is satisfied, we find that  
${\Gamma_2^{\text{dd}(2)}\approx - \frac{\gamma v^2}{E_J^2}\frac{2}{3} 
\left (\gamma \tau \right )^2}$. Since this represents a negligible term, 
the decay of the Cooper pair box is effectively dominated 
by the decay rate $\Gamma_2^{\text{dd}(1)}$.

\begin{figure}[h]
\psfrag{x}{$\tau$}
\psfrag{y}{$\Gamma_2^{\text{dd}}/\Gamma_2$}
\psfrag{a}{$\hspace{-3.8mm}\Gamma_2^{\text{dd}(1)}$}
\psfrag{b}{$\Gamma_2^{\text{dd}(2)}$}
\includegraphics[width=8cm]{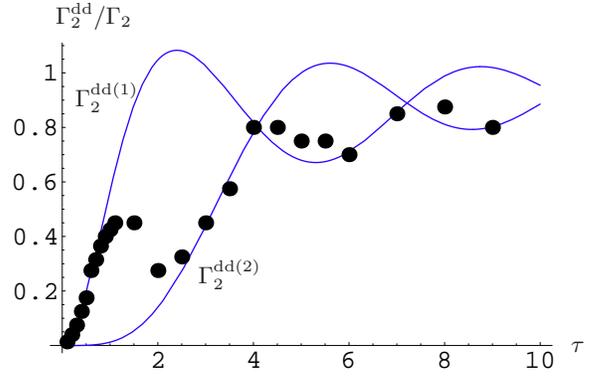}
\caption{
Weak fluctuator ($E_J=1$, $v=0.1$, $\gamma=1$).
The lines show the Eq. (\ref{y1}) and Eq. (\ref{y2}),
points are numerical simulations. From Eq. (\ref{y1}) we have 
$\Gamma_2=\gamma v^2/(E_J^2+4\gamma^2)$.
\label{optGaussY}}
\end{figure}

As an illustrative example, in Fig.~\ref{optGaussY} we consider the 
dissipative dynamics of a qubit that is coupled to a fluctuator in 
this regime and we compare the exact solutions given in Eq. (\ref{y1}) and 
Eq. (\ref{y2})  to the numerics. Notice that for small time interval $\tau$ 
the dissipative dynamics is governed  by the
larger decay rate, presumably because larger weight is given to this solution. 

\begin{figure}[h]
\psfrag{x}{$\tau$}
\psfrag{y}{$\Gamma_2^{\text{dd}}/\Gamma_2$}
\psfrag{u}{$\hspace*{4mm}\small{\tau}$}
\psfrag{v}{$\vspace{2mm}\small{\Gamma_2^{\text{dd}(1)}/\Gamma_2}$}
\psfrag{a}{$\Gamma_2^{\text{dd}(1)}$}
\psfrag{b}{$\Gamma_2^{\text{dd}(2)}$}
\includegraphics[width=8cm]{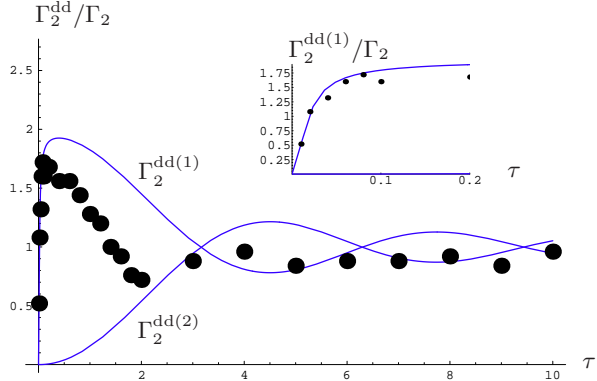}
\caption{
Weak fluctuator ($E_J=1$, $v=0.1$, $\gamma=100$). The lines
illustrate Eq.~(\ref{y1}) and Eq.~(\ref{y2}) and the black points represent
the numerical results.
The inset shows the analytic and numerical solutions for small values of $\tau$. 
From Eq. (\ref{w12}) we have ${\Gamma_2=v^2/4\gamma}$.
\label{optGaussYfast}}
\end{figure}

(ii) {\em weak} fluctuator with ${\gamma\gg E_J}$. \\In this limit, Eq. (\ref{y1}) and Eq. (\ref{y2}) become:
\begin{eqnarray}
\Gamma_2^{\text{dd}(1)} &\approx& 
  \frac{v^2}{4\gamma}\left[1+\frac{\sin\left (E_J \tau \right )}{E_J \tau}
 - 2\cos^2\left (\frac{E_J\tau}{2} \right )\frac{\tanh \left (\gamma\tau \right )}{\gamma\tau}\right] \notag \\ 
\Gamma_2^{\text{dd}(2)} &\approx&
\frac{v^2}{4\gamma}\left[1-\frac{\sin\left (E_J \tau \right )}{E_J \tau}\right]
\label{w12}
\end{eqnarray}
The efficiency of the dynamical decoupling can be easily inferred from the 
decaying rate $\Gamma_2^{\text{dd}(2)}$ given in Eq. (\ref{w12}) and, as 
in the previous case, the condition to satisfy is: ${\tau \le \pi/E_J}$.
By expanding in the limit ${E_J \tau \ll 1}$ we obtain that 
${\Gamma_2^{\text{dd}(1)} \approx \frac{v^2}{4\gamma} \left ( 2-\frac{(E_J \tau)^2}{6} \right ) \left ( \frac{\tanh \left (\gamma \tau \right )}{\gamma \tau} - 1 \right )}$. It follows that 
in order to keep the latter negligible one 
has also to satisfy the condition ${\tau \ll 1/\gamma}$. As a result, in 
this limit, one should choose as condition for the efficient decoupling 
the strongest one, i.e. ${\tau \ll 1/\gamma}$. 
This behavior is illustrated in Fig \ref{optGaussYfast}. 

\begin{figure}[h]
\psfrag{x}{$\tau$}
\psfrag{y}{$\Gamma_2^{\text{dd}}/\Gamma_2$}
\psfrag{a}{$\hspace{-3.5mm}\Gamma_2^{\text{dd}(1)}$}
\psfrag{b}{$\Gamma_2^{\text{dd}(2)}$}
\includegraphics[width=8cm]{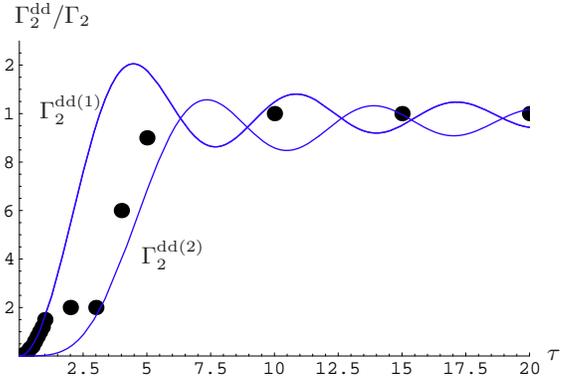}
\caption{Strong fluctuator ($E_J=1$, $v=0.1$, $\gamma=0.01$). The solid lines
show Eq.~(\ref{y1}) and Eq.~(\ref{y2}) and the black points illustrate numerics.
From Eq.~(\ref{sss}) we have $\Gamma_2=\gamma v^2/E_J^2$.
\label{optNonGaussY}}
\end{figure}

Let us now consider the case of a strong fluctuator coupled to the qubit.
The dephasing rates read now:
\begin{widetext}
\begin{eqnarray} 
  \Gamma_2^{\text{dd}(1)} &=&\frac{\gamma v^2}{\Delta^2}
   \left[1-\frac{ \sin \left (\tau \Delta \right )}{\tau \Delta}\right] \label{sss} \\
  \Gamma_2^{\text{dd}(2)} &=& \frac{\gamma E_J^2}{\Delta^2}
   \left [ \frac{2 \Delta^2 + v^2 \left [ 2 \cos (\tau \Delta) + 1 -
3 \frac{\sin (\tau \Delta )}{\tau \Delta} \right ]}{\Delta^2+E_J^2+v^2 \cos(\tau \Delta)} -
\sqrt{\frac{4 v^2 \left [\cot \left (\frac{\tau \Delta}{2} \right )-
\frac{2}{\tau \Delta} \right ]^2}{E_J^2+\Delta ^2 \cot^2 \left (\frac{\tau \Delta}{2}\right )}+ \left [ \frac{2 E_J^2 - v^2 \left ( 1-3 \frac{\sin \left (\tau \Delta \right )}{\tau \Delta} \right )}{\Delta^2+E_J^2+v^2 \cos(\tau \Delta)}\right ]^2}  \right ]
\notag
\end{eqnarray}
\end{widetext}
where $\Delta = \sqrt{E_J^2+v^2}$.
Since we are only considering noise that is weak compared to the level 
splitting of the qubit, ${v\ll E_J}$, and the fluctuator is strong, i.e. 
${\gamma<v}$, only the case ${\gamma \ll E_J}$ is relevant. 
In order to have efficient dynamical decoupling of $\Gamma_2^{\text{dd}(1)}$
it is immediate to realize that we need to set
${\tau \le \pi/E_J}$. Once this constrained is satisfied, we find that
$\Gamma_2^{\text{dd}(2)}  \approx 0$. Therefore the dissipative 
dynamics is finally dominated by $\Gamma_2^{\text{dd}(1)}$.
An example is shown in Fig. \ref{optNonGaussY}.

\section{Imperfect pulses.}
\label{imperfect}
So far we have assumed that the control pulses are ideal. Namely, they 
are zero-width $\pi$ pulses applied exactly along the $x$ or $y$ axis 
of the Boch sphere.
However, in real applications it is plausible that 
the duration of the pulses exhibits random fluctuations, leading 
to error in the rotation angle 
and that the effective rotation axis might be  deviated from the 
$x$ or $y$ axis. 

Clearly, if the pulses are imperfect and the repetition rate between 
the control cycles is  high, there is the possibility that 
the dynamical decoupling introduces more dephasing than what it is 
effectively capable to remove. As a result, given the noise characteristics 
of both the environment and the control pulses, there must exist an 
optimal pulse rate that sets the best decoupling performance. This 
section aims to determine this optimal pulse rate.

Control sequences with
imperfect pulses have been already considered in two recent works 
\cite{Khodjasteh2005,Gutmann2004}. Our analysis differs from the 
previous studies because: (i) it takes into account the 
qubit dynamics between the control pulses; (ii) by using
the analytic solutions derived in the previous 
sections, we are able to provide bounds on the amount of noise 
that can be tolerated in the pulses while still achieving efficient 
decoupling. 

In order to calculate the optimal pulse rate it is convenient to proceed by 
steps. First we study the dissipative 
dynamics for an ideal qubit (without charge noise) that is dynamically 
decoupled with control sequences of noisy pulses, then 
we extend this analysis to a qubit afflicted by a charge fluctuator.

\subsection{Dephasing due only to errors in the pulses}
\label{noise}
In this section we study the dissipative dynamics of an ideal qubit that 
is subject to dynamical decoupling with noise in the $\pi$ pulses.
Specifically, we  assume that the effective 
rotation realized by each pulse fluctuates around the 
expected values: i.e. ${\theta=\pi \pm \delta}$, where 
$\delta$ is a random variable.

In the absence of a fluctuator
the time evolution
of the vector state on the Bloch sphere from the end of one pulse to the end of next pulse 
is then given by 
\begin{eqnarray}
U_i&=&e^{(\pi + \delta_i) R_x}e^{\delta E_c \tau R_z} \qquad \mbox{at pure dephasing} \; \notag \\
U_i&=&e^{(\pi + \delta_i) R_y}e^{E_J \tau R_x} \qquad \mbox{at optimal point}\; \notag
\end{eqnarray}
After $N$ pulses, we find that the total time evolution 
reads: ${U=\prod_{i=1}^N U_i}$.  
Let us now assume that the fluctuations $\delta_i$ (${i=1,\dots N}$) 
are uncorrelated and that they are Gaussian distributed. 
The average time evolution can then be easily calculated 
as $\langle U\rangle=\prod_i\langle U_i\rangle=\langle U_i\rangle^N$ and
the decay rate of the qubit can be inferred by evaluating the 
eigenvalues of $\langle U_i \rangle$. Both for the case of pure dephasing and at the optimal point we find that, in the limit $\langle\delta^2\rangle\ll1$, 
the eigenvalues are:
\begin{eqnarray}
 \alpha_0 &=& -p_c, \notag \\ 
\alpha_\pm &=& \frac{1}{2}\cos\left (\Delta\tau \right )(1-p_c)
   \pm\sqrt{\frac{1}{4}\cos^2\left (\Delta\tau \right )(1-p_c)^2+p_c}
\notag
\end{eqnarray}
where ${p_c=\langle\cos\delta\rangle}$ and ${\Delta=\delta E_c}$ at
pure dephasing and ${\Delta = E_J}$ at qubit optimal point.

It follows that, to the lowest order in  ${\langle\delta^2\rangle}$, 
the dissipative dynamic of the qubit is controlled by
three possible decaying rates ${\Gamma_2^{\text{dd}(i)}=-\ln|\alpha_i|/\tau}$ for 
${i=0,\pm}$. Precisely we find that:
\begin{eqnarray}
 \Gamma_2^{\text{dd}(0)}&=& \frac{\langle\delta^2\rangle}{2\tau} \label{uno1} \\ 
 \Gamma_2^{\text{dd}(\pm)}&=& \frac{\langle\delta^2\rangle}{4\tau}(1\pm\cos\Delta\tau)
\label{due2}
\end{eqnarray}
At long times we expect that the qubit decay is controlled by the 
largest dephasing rate given in Eq.~(\ref{uno1}).

\begin{figure}[h]
\psfrag{u}{$\tau$}
\psfrag{v}{$A_\pm$}
\psfrag{x}{$\tau$}
\psfrag{y}{$\Gamma_2^{\text{dd}}/\Gamma_2$}
\includegraphics[width=8cm]{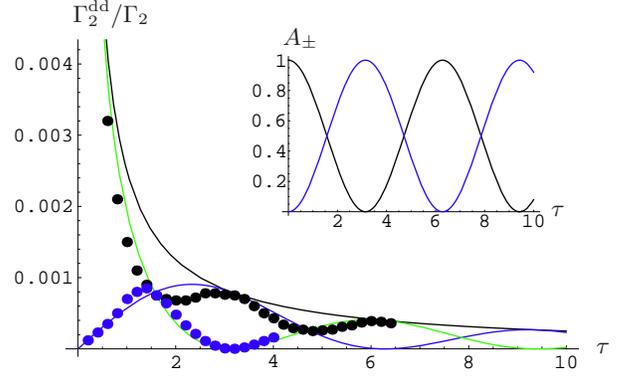}
\caption{
The three decay rates $\Gamma_2^{\text{dd}(0)}$ (black), 
$\Gamma_2^{\text{dd}(-)}$ (blue), $\Gamma_2^{\text{dd}(+)}$ (green)
 ($\delta$ uniformly distributed 
between -0.12 and 0.12 and $\Delta=1$). The black (blue) 
points are the result of 
numerical simulation, with the initial state on the $y$-axis ($x$-axis)
 of the Bloch sphere and pulses around the $x$-axis.  
The inset shows the weights $A_+$ (black) and $A_-$ (blue) for the 
initial state  on the $y$-axis. If the initial state is on the $x$-axis 
the weights interchange.
\label{f4}}
\end{figure}

At small times, the controlled dissipative dynamics is more complicated.
In order to understand it,
we consider the case of a qubit that is tuned at pure dephasing ${\Delta E= E_J=1}$
and it is subject to sequences of $\pi_x$ pulses with fluctuating noise 
${-0.12 \le \delta \le 0.12}$ uniformly distributed. We solve numerically the
dissipative dynamics for two different initial preparation of the qubit: 
either the qubit initial state is prepared along the $x$-axis or the $y$-axis of 
the Bloch sphere.  The results of our simulations are illustrated in 
Fig.~\ref{f4}.

Notice that, depending on the initial conditions, the dissipative
dynamics is dominated by different dephasing rates. 
For a qubit initially prepared along the $y$-axis ($x$-axis)
of the Bloch sphere the dynamics is dominated by the dephasing rate 
${\Gamma_2^{\text{dd}{+}}}$ (${\Gamma_2^{\text{dd}{-}}}$). In order
to understand this, we need to evaluate the weights of the 
different components.

Let us indicate with 
${\bf s_0}$ the initial qubit vector state. It is convenient to write it 
as a linear combination of the eigenvectors ${{\bf v}_i \equiv (v_i^0,v_i^+,v_i^-)}$ of the average operator $\langle U_i\rangle$ describing the evolution of the qubit vector state on the Bloch sphere during a control cycle, i.e.
${{\bf s}_0=\sum_\kappa A_\kappa v_i^\kappa}$, with ${\kappa=0,+,-}$.
After $N$ control cycles, the qubit vector state reads: 
${{\bf s}= \langle U_i \rangle^N {\bf s}_0= \sum_\kappa A_\kappa \alpha_\kappa^N v_i^\kappa}$ and the coefficient $A_i$ can be easily found.

In the considered example, when the qubit is initially prepared along the $y$-axis of the Bloch sphere, i.e. ${\bf s}_0={\bf e}_y$, we find that the coefficients reads: 
$A_0=0$ and $A_\pm = (1\pm\cos\Delta\tau)/2$. Therefore we understand 
immediately why in Fig.~\ref{f4} the dephasing rate $\Gamma_2^{\text{dd}(0)}$
never dominates the dissipative dynamics. Moreover it is clear that, according to the 
oscillatory behavior of the coefficient $A_\pm$ displayed in the inset, the dissipative qubit 
dynamics at different time intervals $\tau$ between the control 
pulses is alternatively dominated by the two dephasing rates ${\Gamma_2^{\text{dd}\pm}}$.

Notice that, in a realistic situation where dynamical decoupling sequences 
are introduced in the protocol during the computation, we might not know
exactly which is the effective state of the qubit. As a result, from the point of view of the efficiency of the control when fluctuating noise in 
the pulses is considered, it is 
reasonable to assume that the dissipation is always given by 
the largest decay rate ${\Gamma_2^{\text{dd}(0)}}$ at all interval times $\tau$
between the pulses.

\subsection{Dephasing due to error in the pulses and fluctuator noise}
\label{ci}
Let us now consider that during the dynamical decoupling there is both 
noise coming from the pulses and a charge fluctuator that is coupled to 
the Cooper pair box.
Since the two sources of noise are independent, we expect that the 
relaxation rates will just add. We could not prove this conjecture 
analytically, but results of numerical 
simulations clearly demonstrate that this is always the case. 

\begin{figure}[h]
\psfrag{x}{$\tau$}
\psfrag{y}{$\Gamma_2^{\text{dd}}/\Gamma_2$}
\includegraphics[width=8cm]{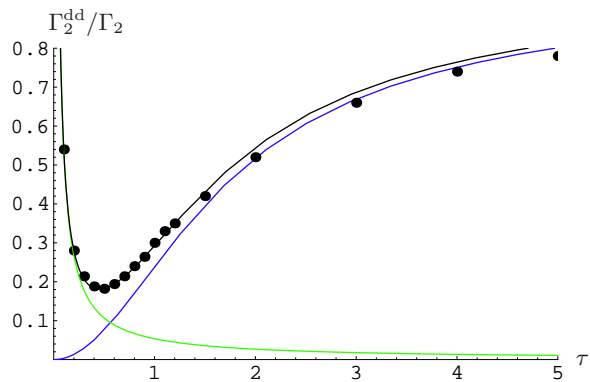}
\caption{$\Gamma_2$ with $\delta$ uniformly distributed between -0.04 and 0.04
and $\Delta=1$, $E_J=0$. One weak fluctuator, $v=0.1$, $\gamma=1$.
Green - noise only in the pulses, blue - noise only from the fluctuator, 
black - the sum of the two noise contributions. Black points are 
numerical simulation.  
From Eq.~(\ref{bbwl}) we have $\Gamma_2=v^2/2\gamma$.
\label{f6}}
\end{figure}

As an example, in Fig.~\ref{f6} we illustrate the
dissipative dynamics of a qubit that is coupled to a weak fluctuator 
at pure dephasing.
 In the figure we plot: 
the analytic solution for the dissipative dynamics when there is noise 
only in the pulses given in Eq. (\ref{uno1}) (green), 
the analytic solution for the dephasing rate due to the interaction 
of a weak fluctuator 
derived in Eq. (\ref{bbwl}) (blue) and the total dephasing rate 
resulting from the sum of the two (black). 
Notice that there is a good agreement between numerics and analytic results.  

By looking at Fig.~\ref{f6} we see that, by 
decreasing the time interval $\tau$ between the pulses, the dephasing rate
initially decreases, then at a particular value of $\tau$ it reaches 
a pronounced minimum where the best dynamical decoupling performance is 
achieved and finally, by further decreasing $\tau$, the performance of the 
decoupling deteriorates. 
Since the efficiency of the
decoupling (i.e. the value of the minimum) is related to the level 
of noise in the pulses, it might be interesting to
estimate how much noise in the pulses can be tolerated in order to have
an efficient suppression of the noise due to the fluctuator. 

To this aim, one has to find the time interval 
$\tau$ between the pulses when dynamical decoupling starts to reduce the decoherence rate.
If the decoherence rate because of the noise on the flips 
($\sim\langle\delta^2\rangle/\tau$) at this pulse rate 
is small compared to the one from fluctuators (which will be of the order of 
the rate without dynamical decoupling) there will be a range of $\tau$ below 
this point where dynamical decoupling is working, and it will not be 
destroyed by the pulse noise. 

In Sec.~\ref{ideal} we have demonstrated that, 
depending on the nature of the fluctuator, efficient dynamical decoupling 
is achieved for ${\tau\lesssim1/\gamma}$ if the fluctuator is weakly coupled 
to the qubit and for ${\tau \lesssim 1/v}$ if the fluctuator is non-Gaussian.
As a consequence, depending on the type of fluctuator interacting with 
the qubit, we find that the control sequences are efficient when the noise 
in the pulses satisfy the following conditions:
\begin{eqnarray}
\langle\delta^2\rangle &<& \left (\frac{v}{\gamma} \right )^2 \qquad ~~~~ \text{for weak fluctuator} \notag \\
\langle\delta^2\rangle &<& \frac{\gamma}{v} \qquad ~~~~~~~~~~ \text{for strong fluctuator} \notag 
\end{eqnarray}
Notice that the noise tolerated in the pulses 
is bounded by the second power of the ratio 
${v/\gamma <1}$ for a weak fluctuator and only by the first power 
of ${\gamma/v<1}$ for a non Gaussian fluctuator. Therefore, in a qubit afflicted by 
a strong fluctuator, efficient dynamical decoupling 
can be achieved with higher noise in the pulses.

A similar analysis can be done for a qubit that is tuned at the optimal point.
Using the analytic solutions given in Sec.~\ref{idealop} we find 
that an efficient dynamical decoupling can be achieved
\begin{itemize}
\item[-]For weak fluctuator, i.e. $v<\gamma$, when:
\[
\left \{ \begin{array}{ll}
\langle\delta^2\rangle < \frac{\gamma v^2}{E_J^3} & \mbox{ if $\gamma\ll E_J$};\\
\langle\delta^2\rangle < \frac{v^2}{\gamma^2} & \mbox{if $\gamma\gg E_J$. }
\end{array} \right . \]

\item[-] For strong fluctuator, i.e $v>\gamma$, when:
\[
\langle\delta^2\rangle < \frac{\gamma v^2}{E_J^3}
\]
\end{itemize}

\section{Conclusions.}
\label{conc}
In this paper we studied the dissipative dynamics 
of a Cooper pair box (qubit) afflicted by charge noise that is 
dynamically decoupled. By focusing on the case where a single fluctuator 
is coupled to the qubit, we investigated the efficiency of ideal 
(without errors in the pulses) decoupling sequences. At qubit pure dephasing, irrespective of whether one applies sequences of
$\pi_x$ or $\pi_y$ pulses, we found that the efficiency of the decoupling 
is strongly dependent on the characteristics of the fluctuator. 
For the case of a weak (Gaussian) fluctuator an efficient suppression of the 
noise is achieved by choosing the time interval between the pulse less than 
the switching rate of the fluctuator, while
 for a strong (non-Gaussian)
 fluctuator the relevant decoupling rate is determined by the coupling 
strength of the fluctuator to the qubit. As a result, in this limit dynamical decoupling techniques can be successfully implemented as a diagnostic tool to infer spectral informations on the nature of the fluctuator coupled to the qubit.

A different situation is encountered at the qubit optimal point, 
where we showed 
that the decoupling performance varies considerably depending whether 
control sequences of $\pi_x$ or $\pi_y$ pulses are used to decouple the qubit.
Both for the case of weak and strong fluctuator we found that the 
dissipative dynamics is dominated by the qubit energy scale. Although for the special case of weak fluctuator characterized by a switching rate greater 
than the qubit energy level splitting we observed that the efficiency of the dynamical decoupling is achieved by choosing time interval between the pulses less than the switching rate of the fluctuator.

We have extended this analysis in order to include also errors in the pulses.
We showed that, depending on the characteristic of the fluctuator and the 
error in the pulses, there must exist an optimal pulse rate that sets the best decoupling performance. Using the analytic solutions for the dissipative qubit dynamics derived both at pure dephasing and at the optimal point, we found upper bounds on  the level of noise present in the pulses that can be tolerated while still achieving an efficient dynamical decoupling.

We believe that an interesting future research direction is to extend this 
analysis in order to take into account the quantum nature of the fluctuator.
It would be extremely interesting to see whether dynamical decoupling techniques could be useful in order to understand the microscopic mechanisms of 
relaxations at low Temperature of the {\em quantum} Two Level Systems present in the 
amorphous materials of the barrier and substrate.    

\acknowledgments 
This work was supported by the National Security Agency (NSA) under Army Research Office (ARO) contract number W911NF-06-1-0208 and the Norwegian Research 
Council via a StorForsk program. 

\vspace*{-2mm}

\bibliography{DecfinL.bbl}

\begin{thebibliography}{44}
\bibitem[{Joa()}]{Joaknote}
\bibinfo{note}{In a RTP the switchings are independent and the probability
  $P_n$ of $n$ switching events in a certain time $t$ is given by Poisson
  statistics, ${P_n(t) = (\gamma_it)^ne^{-\gamma_it}/n!}$. The fact that we
  assume a symmetric RTP means that the fluctuator will spend an equal amount
  of time in both states. This assumption can easily be relaxed, but we believe
  it is sufficient for our purposes.}
\end{thebibliography}
\end{document}